\documentclass[a4paper,11pt]{article}
\pdfoutput=1 

\usepackage{jcappub} 
\usepackage[T1]{fontenc} 
\usepackage{subcaption}
\usepackage{booktabs} 
\usepackage{graphicx}
\graphicspath{{used_figures/}} 
\usepackage{times}
\usepackage{textcomp}

\title{Measurement of Cosmic Muon angular distribution and vertical integrated flux by 
  2\,m\,$\times$\,2\,m\, RPC stack at IICHEP-Madurai}

\author[a,b,1]{S. Pethuraj}
\author[b]{V.M. Datar}
\author[b]{G. Majumder}
\author[c]{N.K. Mondal}
\author[b]{K.C. Ravindran}
\author[b]{B. Satyanarayana}

\affiliation[a]{Homi Bhabha National Institute,\\ Mumbai-400094}
\affiliation[b]{Tata Institute of Fundamental Research,\\ Mumbai-400005}
\affiliation[c]{Saha Institute of Nuclear Physics,\\ Kolkata-700064} 

\emailAdd{s.pethuraj@tifr.res.in}
\emailAdd{vivek.datar@tifr.res.in}
\emailAdd{gobinda@tifr.res.in}
\emailAdd{nabak.mondal@gmail.com} 
\emailAdd{kcravi@tifr.res.in}
\emailAdd{bsn@tifr.res.in}

\abstract{ The 50 \,kton\, INO-ICAL is a proposed underground high energy physics experiment at Theni,
  India ($9^{\circ}57'N$, $77^{\circ}16'E$) to study the 
  neutrino oscillation parameters using atmospheric neutrinos. The Resistive Plate Chamber
  (RPC) has been chosen as the active detector element for the ICAL detector.
  An experimental setup consisting of 12 layers of glass RPCs of size 2\,m\,$\times$\,2\,m has been built at IICHEP, Madurai to 
  study the long term stability and performance of RPCs which are produced on a large scale in Indian 
  industry. In this paper, the studies on the performance of RPCs are presented along with the 
  angular distribution of muons at Madurai ($9^{\circ}56'N,78^{\circ}00'E$ 
  and  Altitude $\approx$\,160\,m from sea level).}

\keywords{cosmic ray experiments, cosmic rays detectors}

\begin{document} 
\maketitle 
\flushbottom

\section{Introduction}
\label{sec:intro}

The primary cosmic rays originating from outer space
mostly consist of high energy protons and a smaller fraction of other high 
Z-nuclei. The interaction of primary cosmic rays with the air molecules 
in the earth's atmosphere results in showers of secondary particles, which are mostly composed of
pions ($\pi^{+},\pi^{-}$ and $\pi^{0}$),  in the upper 
atmosphere. The neutral pion mainly decays via electro-magnetic interaction
and produces $2\gamma$ while charged pions decay via the weak interaction leading to muons and neutrinos, $\pi^{+} \rightarrow \mu^{+} + \nu_{\mu}$ and $\pi^{-} \rightarrow \mu^{-} + \overline{\nu_{\mu}}$. Muons then decay through, $\mu^{+} \rightarrow e^{+} + \overline{\nu_{\mu}} + \nu_{e}$ and $\mu^{-} \rightarrow e^{-} + \nu_{\mu} + \overline{\nu_{e}}$. INO will be an underground experiment to precisely measure neutrino oscillation parameters using signature of neutrino interactions. It will also determine the sign of the 2-3 mass-squared difference, $\Delta m^2_{32}$ (= $m^2_3 - m^2_2$) through matter effects, the value of the leptonic CP phase and, last but not the least, the search for any non-standard effect beyond neutrino oscillations \cite{inowhite}. About 28000 glass Resistive Plate Chamber (RPC) \cite{santonico} of size $\sim$\,(2\,$\times$\,2\,m$^2$) will be used as sensitive detectors to measure energy and direction of neutrinos.

At the sea level muons are the most abundant charged particles in cosmic rays. Most 
muons are produced high in the atmosphere (typically at altitude of about 15 km) and lose energy of about 
2 GeV via ionization before reaching the ground. The primary cosmic rays are 
more or less isotropic but the observed muon angular 
distribution at various altitudes and latitudes follows the expression 
$I_{\theta} = I_{0}cos^{n}\theta$, where $I_{0}$ is the vertical integrated muon 
flux($cm^{-2} s^{-1} sr^{-1}$) and $\theta$ is the zenith angle of muon. The exponent, 
n and $I_{0}$ depend on momentum cut off, latitude and altitude. The 
experimentally observed and Monte-Carlo (GEANT4) generated 
theta distributions are used to extract the exponent of the 
$cos\theta$ distribution in data at Madurai, India. The vertical flux is estimated 
from the integral of the theta distribution and normalised by selection efficiency, trigger 
efficiency, DAQ efficiency, solid angle coverage and integrated time for data collection.
All detector parameters like efficiencies, multiplicities and position 
smearing are obtained from observed data, which are used in the Monte-Carlo event generation. 
This muon flux can be used as a input of neutrino generator for better neutrino flux at the INO experiment.

Section \ref{sec:setup} describes the experimental setup followed by the preliminary analysis of data 
has been described in Section \ref{sec:data}. The MC simulation framework using GEANT4 toolkit is explained in 
Section \ref{sec:mc}. The techniques used to estimate the muon flux is
presented in Section \ref{sec:expint}. The systematic uncertainties in the measurement and the 
results are discussed in Section \ref{sec:syst} and \ref{sec:compr} respectively. The comparison of the present
results with the previous measurements at different geomagnetic cut off is also shown in Section \ref{sec:compr}.
In Section \ref{sec:conclu}, the major results of current study are summarized and future prospects are discussed.

\section{The 2m x 2m Detector Setup }
\label{sec:setup}
\begin{figure}[h]
  \begin{subfigure}{0.5\textwidth}
    \includegraphics[width=0.9\linewidth, height=5cm]{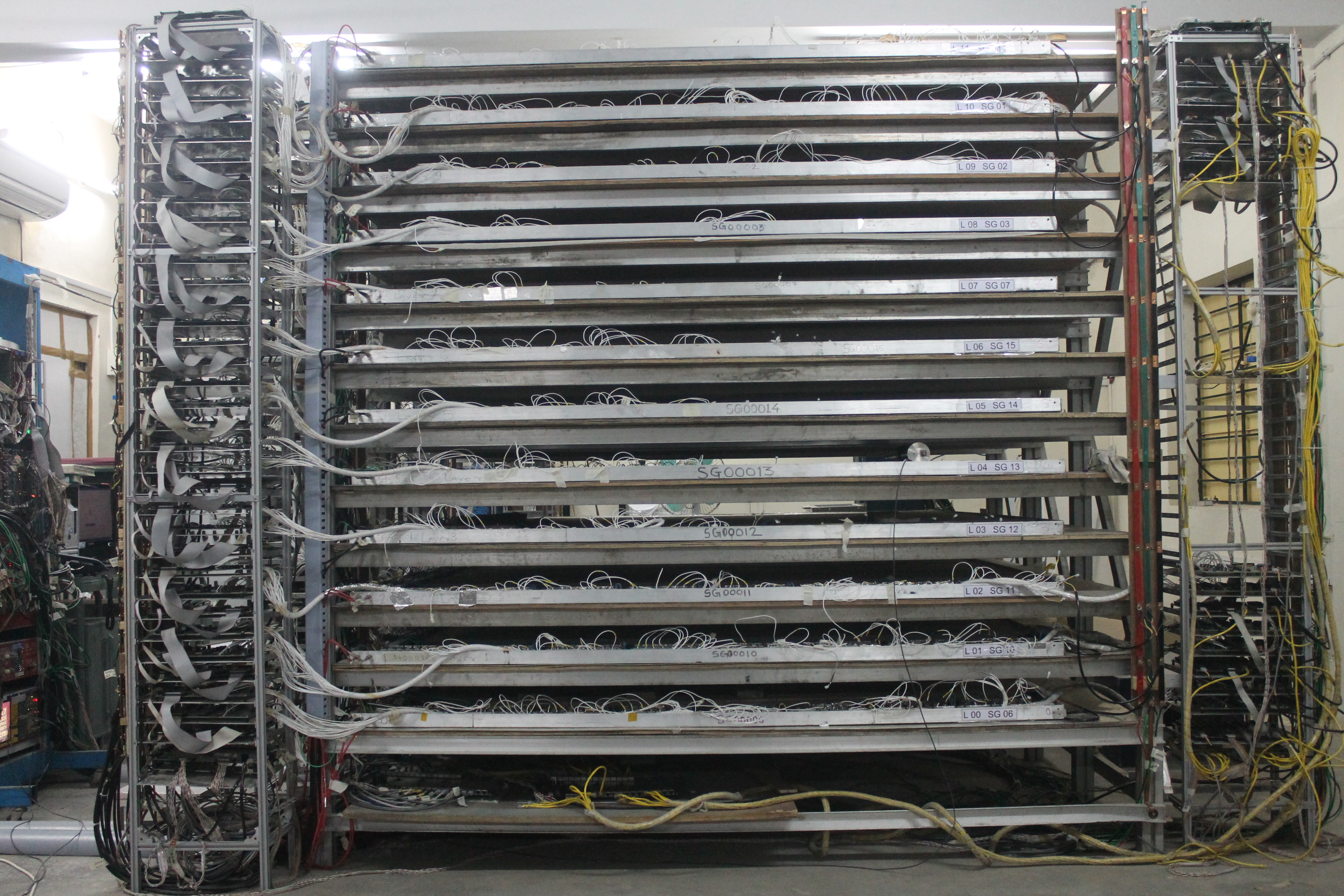} 
    \caption{Detector Setup}
    \label{fig:subim1}
  \end{subfigure}
  \begin{subfigure}{0.5\textwidth}
    \includegraphics[width=0.9\linewidth, height=5cm]{SIGNAL_FLOW_071216.png}
    \caption{Signal Flow }
    \label{fig:subim2}
  \end{subfigure}
  
  \caption{12-layer detector stack of 2m x 2m RPCs and its signal flow.}
  
\end{figure}

The detector stack(shown in figure \ref{fig:subim1}) used in this study consists of 12 layers 
of glass RPCs of size 2\,m\,$\times$\,2\,m\footnote{The size of RPC glasses is 1.84\,m\,$\times$\,1.91\,m.}
with a gap of 16\,cm between two layers. An RPC is a parallel plate chamber made up of two 
glass electrodes of thickness 3\,mm with a gap of 2\,mm between them. The outer sides of the 
chamber are coated with a thin layer of graphite. 
Each RPC is readout by two orthogonal pickup panels, made up of copper, one on either 
side of the chamber labelled as X- and Y-planes. The X- and Y-plane pickup panels are divide
into 60 and 63 strips respectively with the strip width of 2.8\,cm and interstrip gap of 0.2\,cm.
The RPCs are operated in avalanche mode with a non-flammable gas mixture of R134a (95.5\,\%),
iso-butane (4.3\,\%) and $SF_6$ (0.2\,\%), which will be continuously flown through the gas gap.
A differential high voltage of $\pm$\,5\,kV is applied for the operation of RPC.

The flow of signals from the RPCs to the back-end is shown in figure 
\ref{fig:subim2}. The detailed description of signal processing and Data Acquisition system (DAQ) can be found in \cite{elec2}. 
Due to shortage of electronics, only the  top three layers (9, 10 and 11) and 
the bottom four layers (0, 1, 2 and 3) are fully instrumented, while the remaining layers have 
electronics only for 32 strips on both the X- and Y- planes. 
The data from layer 4 is not used in this study because of its different data format.
Timing signals from all the strips from both electronic readout
planes of an RPC are ORed to make a 1-fold time signal for each readout plane 
of a layer. These layer-wise 1-fold time signals from X- and Y- planes, are 
recorded separately, by the TDC (V1190B-CAEN, 100 ps LSB). Usually, a coincidence of these 
1-fold time signals from 4 out of 12 layers is used for triggering. But in the current study
only X-plane time signals are used to generate the cosmic muon trigger.
Here, layers 1, 2, 9 and 10 are used in the trigger criteria in order to obtain larger trajectory and better directionality. 
In this study, approximately fourteen million events are used with an average trigger rate of 
60\,Hz. Assuming the energy loss of 1\,GeV/c muons is $\sim$\,2\,MeV per g cm$^{-2}$, 
the minimum momentum cut off in the vertical direction is about 110 MeV.

\section{Analysis of Experimental data}
\label{sec:data}

\begin{figure}
  \includegraphics[width = 1.\linewidth]{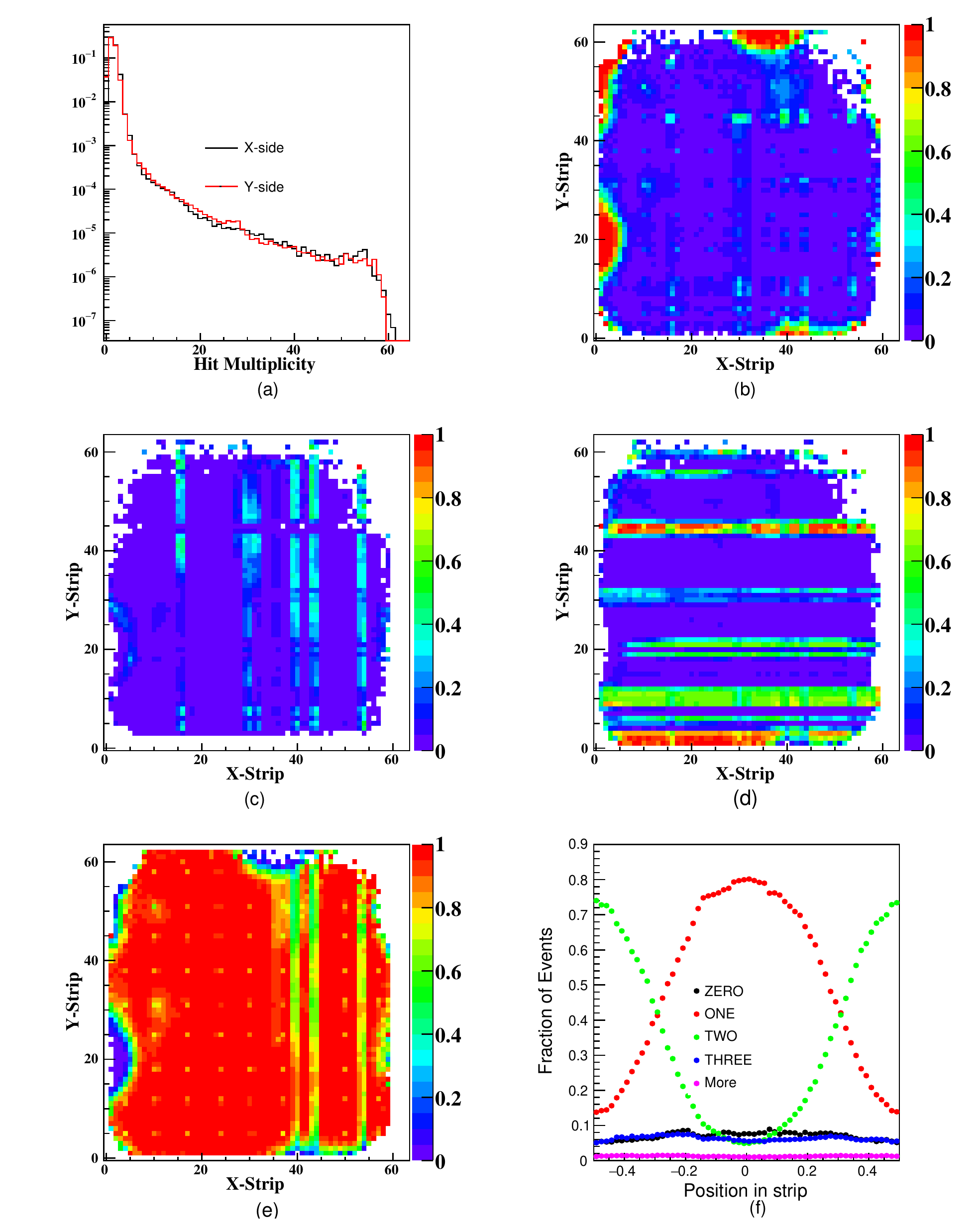}
  \caption{ (a) Strip-hit multiplicity in X-plane and Y-plane in Layer 2., 
    (b) Pixel-wise correlated inefficiencies in Layer 3,
    (c) Pixel-wise uncorrelated inefficiencies in Layer 3 (X-plane), 
    (d) Pixel-wise uncorrelated inefficiencies in Layer 3 (Y-plane), 
    (e) Pixel-wise trigger efficiencies in Layer 3 (X-plane) and 
    (f) strip multiplicity based on muon hit position in strip in Layer 2 (Y-plane). }
  \label{fig:image2}    
\end{figure}

An event data typically consists of hit pattern of strips (one logic bit per
strip) from X- and Y-planes as well as one timing information per plane. This
study is based on the trigger criteria, where at least one strip in the X-plane
of 1st, 2nd, 9th and 10th RPC layers have simultaneous hits within 100 ns time window.
Strip-hit (signal in a strip is more than the discriminator threshold, $V_{th}=$\,-20\,mV) information is used for 
the muon flux measurement. The noise rates of the strips are also used to identify 
any noisy strips for the entire analysis. The average strip multiplicity for the 
hits due to muon trajectory is about 1.5 strips per layer. 
In figure \ref{fig:image2}(a) shows typical strip multiplicities for X- and Y-planes.
It is evident from the figure that a major fraction of events have strip multiplicities one or two.
The events with strip multiplicities beyond three are due to streamers in the RPCs and correlated electronic noise, 
where as zero multiplicity is observed due to inefficiency in an RPC.
During the study, it was also observed that, in a layer for events with strip multiplicity more than three, the position resolution
is of the order of the strip width (3\,cm), which is due to poor localisation of signal. 
For the layers with multiplicity one or two, the observed position resolution was $\sim$\,6\,mm.
Hence, in the present study, the events with at most three consecutive strip hits are considered for analysis.
Using this selection criteria, a noisy layer can be removed without discarding the entire event.
The accepted hit positions in all layers are fitted with a straight 
line in both XZ and YZ planes independently using the equation,
\begin{equation}
  x(/y) = a \times z + b
\end{equation}
where $x$ or $y$ is the hit position (average strip position) from the X- or Y-
plane respectively for Z-th layer,$\texttt{\textquotesingle a\textquotesingle}$ is the slope
and $\texttt{\textquotesingle b\textquotesingle}$is the
intercept. Using these four parameters, the exact
position of the muon trajectory in all the RPC layers can be computed. This is performed 
only when there are at least seven layers with
selected hit position in both the X- and Y-planes for this fit and $\chi^2/ndf$ of the fit is less than 8.

Using the muon data, the physical shifts in the detector position in the X- and Y-planes are 
corrected by an iterative method. In this method one of the layer
is removed in the fitting procedure and the muon trajectory is extrapolated in that layer from the fitted parameters.
The offset is estimated by comparing the measured hit position and extrapolated position in that layer. 
This method is repeated for all the layers. After a few iterations, an overall chamber alignment accuracy of
better than 0.2\,mm can be acheived. A detailed explaination about the selection criteria as well as the alignment
correction procedure can be found in \cite{spal}. Using a similar iterative procedure, multiplicities, 
X- and Y-residues, uncorrelated and correlated inefficiencies and trigger efficiencies for the all the layers are
also calculated.

The hit inefficiencies and trigger efficiencies are calculated for each
3\,cm\,$\times$\,3\,cm pixel, in order to match with strip width.
The algorithm to calculate the pixel wise hit inefficiencies and trigger efficiencies is as follows,

\begin{itemize}
\item The extrapolation error ($\epsilon$) on the hit points in a layer is estimated. The deviation ($\delta$) of a fit point from the mid point of a strip is also calculated. 
\item The trajectories, where $|\delta|$ + $\epsilon$ is within a strip pitch, are only considered.
\item Correlated inefficiencies are estimated using the fraction of events, when a fitted 
  muon has passed through a pixel, but there is no hit in that position in both the X-
  and Y-plane of the detector within 3\,cm of the extrapolated point. The pixel wise correlated inefficiencies for layer 3 are shown in figure \ref{fig:image2}(b).
\item The uncorrelated inefficiencies on X-plane are calculated when the X-plane does not 
  have any hit, but the Y-plane has a hit. So in this case, no hits within 3\,cm of extrapolated point in X-plane but 
  $\Delta$Y (Y-residue) will be less than one strip pitch. Similarly, the uncorrelated inefficiencies on the Y-plane are also calculated.
  The uncorrelated inefficiencies for the X- and Y-plane in layer 3 are shown in figure \ref{fig:image2}(c) and \ref{fig:image2}(d) respectively . 
\item Trigger efficiency is calculated if there is any hit in the layer when a muon has passed through it. A typical trigger efficiency observed in the RPC is shown in figure \ref{fig:image2}(e).
\item The trigger efficiencies of the top six layers are estimated using events triggered by layers 0, 1, 2 and 3. Similarly, the trigger efficiencies of the  bottom six layers are estimated using events triggered by layers 9, 10 and 11.
\end{itemize} 

\section{Monte-Carlo event generation}
\label{sec:mc}
The GEANT4 \cite{geant4} toolkit was used to develop Monte Carlo simulation which incorporated the interaction of muons with the detector medium. The 12 layer RPC stack along with the detector hall was included in the GEANT4 detector geometry to include all the different materials through which a cosmic muon has to traverse. The CORSIKA \cite{heck} software was used to generate the energy spectrum of cosmic muons. In the GEANT4 simulation, the muons are generated above the ceiling of the building with a momentum threshold of 0.11 GeV/c. The various detector parameters like uncorrelated and correlated inefficiencies, trigger efficiencies, strip multiplicity and layer residuals were incorporated during the digitisation process of simulation. The steps followed in the MC event generation are, 
\begin{enumerate}
\item A random position (x,y) is generated in the top most trigger layer (i.e. layer 10). Similarly the zenith angle, $\theta$ is generated using $cos^{2}\theta$ distribution in the range of 0 to 60\,degree\footnote{Due to solid angle coverage of the trigger layers, there is no muon with $\theta \ge 60^\circ$.}. The azimuthal angle, $\phi$ is generated randomly from  $-\pi$ to $\pi$.
\item The generated muon is extrapolated to the bottom trigger layer to test the trigger condition. If the extrapolated position on the bottom trigger layer is inside the RPC detector then, the set of (x,y,$\theta$,$\phi$) are accepted\footnote{This is done to make the computer processing faster.}. The event generation vertex for GEANT4 is calculated for the set of (x,y,$\theta$,$\phi$) and given as input to GEANT4. The energy of the muon is generated from CORSIKA flux and is also given as the input.
\item The simulation of the passage of muons through the detector geometry is performed by the GEANT4. When the muon, passes through an RPC detector volume, the GEANT4 provides the (x,y,z) co-ordinate and the exact time stamp for that point. 
\item The position resolution is used to smear the hit position in each layer. 
\item The smeared coordinates of the hit position are translated into the strip information for the corresponding Z plane.
\item The pixel wise correlated inefficiency map discussed in previous section is used to incorporate the correlated inefficiencies in the simulated event.
\item The position (hit position with respect to the strip center) dependent strip multiplicity, as shown in figure \ref{fig:image2}(f) is used to implement the multiplicity.
\item The uncorrelated inefficiencies for X and Y strips are incorporated independently based on strip multiplicity using the inefficiency map discuss in previous section.
\item The trigger efficiencies are incorporated only for the trigger layers (namely layers 1, 2, 9 and 10) in the X plane to accept an event. 
\item In the experimental data, random noise hits due to electronics and multi-particle shower within the detector volume are also observed. These noise hits are also extracted from data and incorporated duing the digitization process.
\end{enumerate}

\begin{figure}
  \center
  \includegraphics[width=0.9\linewidth]{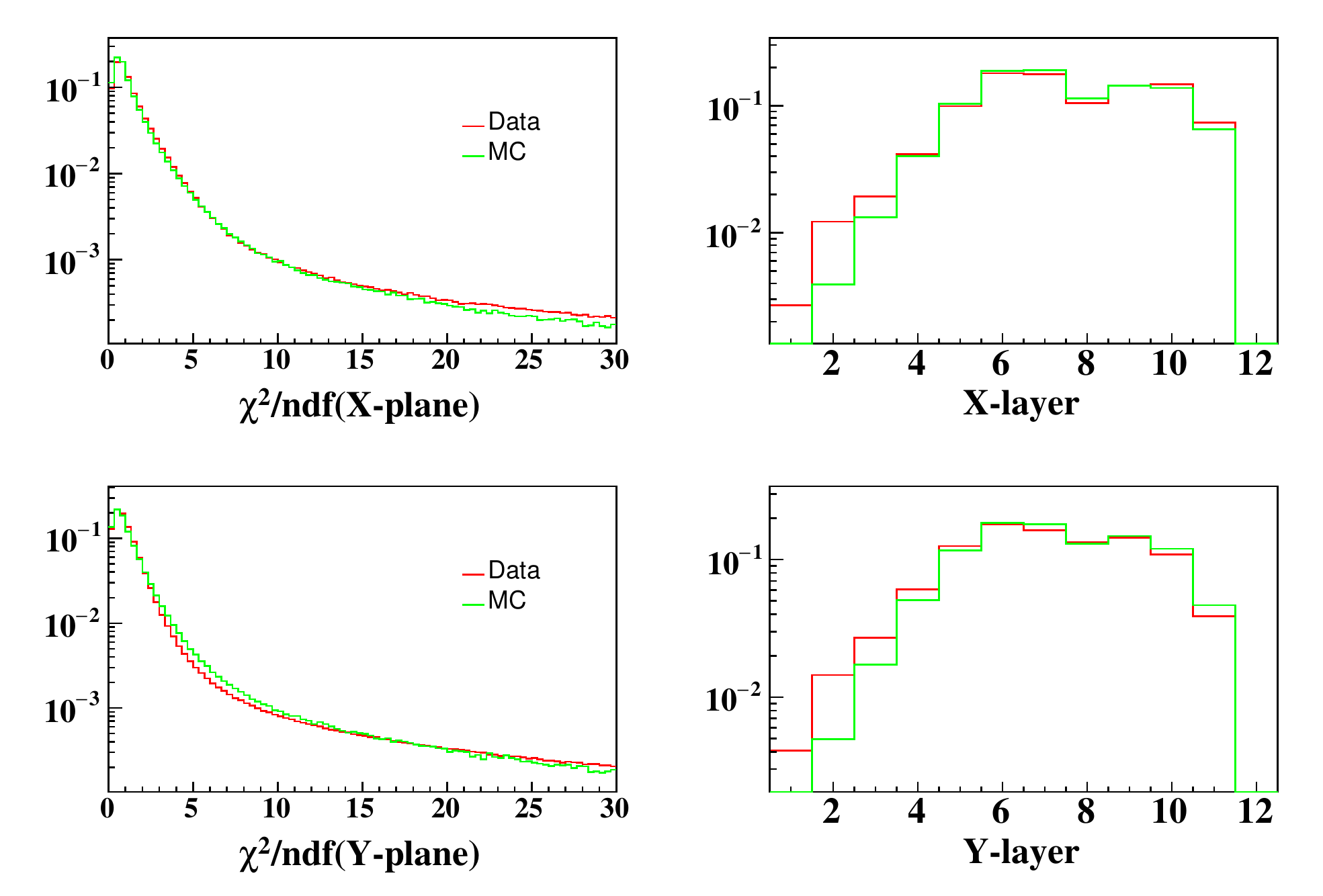} 
  \caption{Comparison of $\chi^{2}/NDF$ and number of used layers in Data and MC.} 
  \label{fig:chindf}
\end{figure}

The simulated events are analysed by the same procedure that is used for the experimental data. The comparison of $\chi^{2}/ndf$  and number of layers hit by muons on both X- and Y- planes is shown in figure \ref{fig:chindf}. A good agreement between data and MC is observed.

\section{Estimation of exponent (n) and vertical flux (I$_{0}$)}
\label{sec:expint}
\subsection{Calculation of Exponent (n)}
Statistically, the best fit value for the exponent (n) can be estimated using the experimentally observed $\theta$ distribution ($N_{Obs}^{\theta_i}$) and the acceptance of muon in the $\theta_i$ bin. The chi-square is defined as,

\begin{equation}
  \label{simpleequation}
  \chi^{2} = \Sigma_{\theta_i} \frac{(N^{\theta_i}_{Obs}- P_{0} \sin\theta_i \cos^{n}\theta_i \ w(\theta_i) )^{2}}{N^{\theta_i}_{Obs}+ (P_{0} \times \sigma_{w})^2},
\end{equation}
where, $\sigma_{w_i}$ is the error on $w(\theta_i)$, $P_0$ is a normalisation constant, a free parameter and $n$ is the exponent.

\begin{figure}[h]
  \begin{center}
    \includegraphics[width=1.\linewidth]{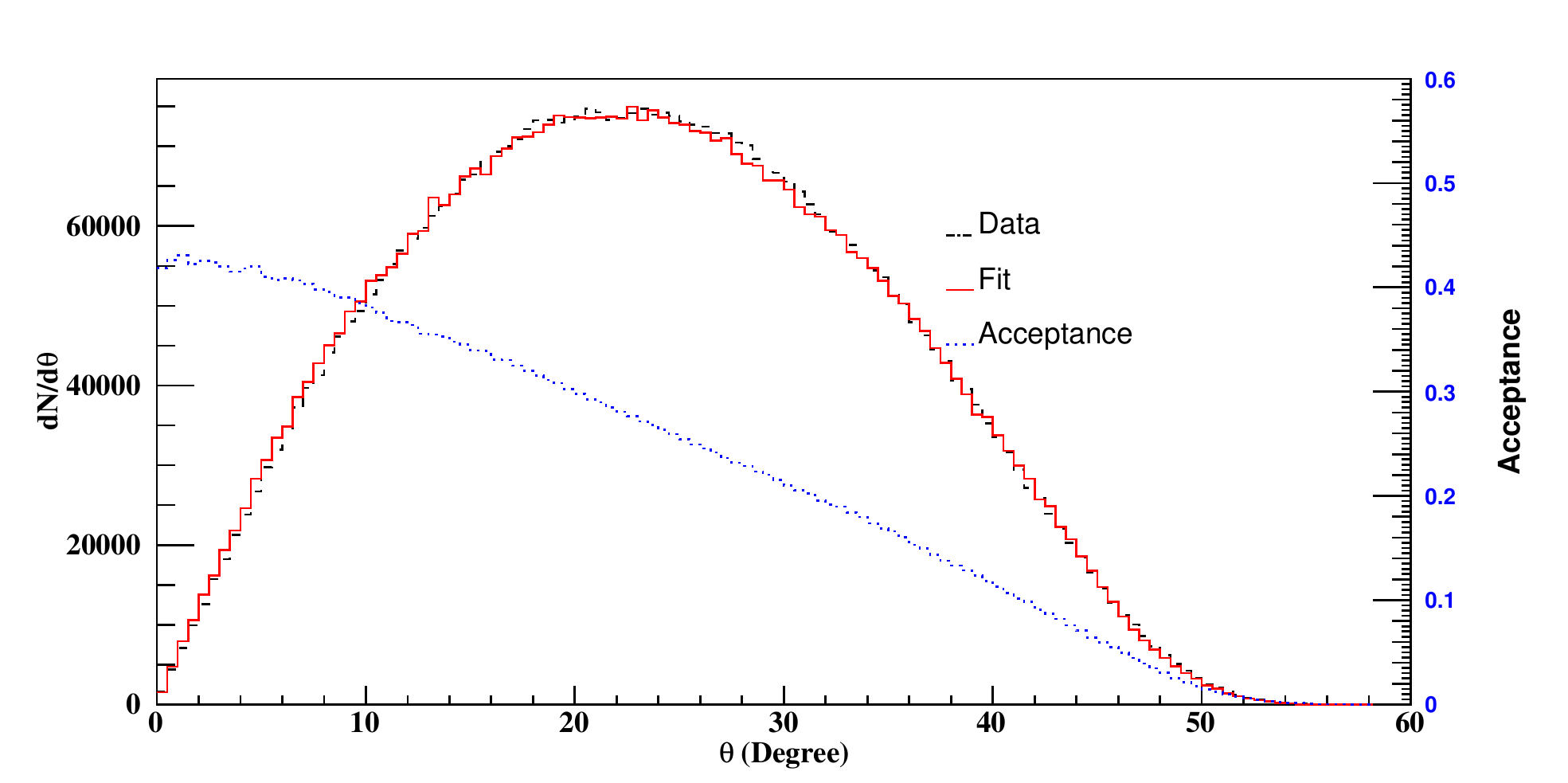}

    \caption{Experimental and fitted theta distributions and acceptance of the detector stack.}
    \label{fig:nvalue}
  \end{center} 
\end{figure}

The comparison of experimentally observed and fitted  $\theta$ distribution for muons which has minimim 7 layers with the hits and the fit $\chi^{2}/ndf < 8$ is shown in figure \ref{fig:nvalue}. The best fit value of exponent is found to be,

\begin{center}
  $n$ = 2.00 $\pm$ 0.04(stat)
\end{center} 

\subsection{Calculation of Integrated Vertical Flux}  

The integral intensity of the vertical muons ($I_0$) can be estimated from the observed $\theta$ distribution which can be given as;
\begin{equation}
  \label{eqn:equation2}
  I_{0} = \frac{I_{data} }{\epsilon_{trig} \times \epsilon_{selec} \times \epsilon_{daq} \times T_{tot} \times \omega }   
\end{equation}
where, $I_{data}$ is the integral of the observed $\theta$ distribution, $\epsilon_{trig}$ is the trigger efficiency, $\epsilon_{selec}$ is the event selection efficiency in data, $\epsilon_{daq}$ is the efficiency due to dead time in the data acquisition system, $T_{tot}$ is the total time taken to record the data(in seconds) including DAQ's dead time (4 ms/event) and $\omega$ is the accepted solid angle times the surface area, which is further defined as,
\begin{equation}
  \label{eqn:equation3}
  \omega = \frac{AN}{N^{\prime}}\int_{0}^{\pi/3} cos^{n}\theta sin\theta d\theta \times 2\,\pi
\end{equation}
where, A is the surface area of the RPC on top triggered layer, N is the number of events accepted when the generated position on the top and bottom trigger layer are inside the detector, $N^{\prime}$ is the  Number of events generated on top trigger layer.
The integral intensity of the vertical muons ($I_0$) with the same selection criteria is found to be,
\begin{center}
  $I_{0}$ = (7.0069 $\pm$ 0.0018(stat)) $\times$ 10$^{-3}$  $cm^{-2} s^{-1} sr^{-1}$
\end{center}


\section{Systematic Uncertainty}
\label{sec:syst}

Besides the statistical errors due to limited number of events in data and MC, 
there are sources of systematic uncertainties in these parameters ($n$ and $I_0$). 
The change in the central value of $n$ and $I_0$ due to uncertainties of different parameters listed below and corresponding change in $I_0$ and $n$ 
along with the change in percentage with respect to default value are given in table \ref{tab:nvalue}.  
\begin{enumerate}
\item In the calculations, a minimum of 7 layers in both X- and Y-plane fit are used 
  as a trade off between the fitting quality and the total statistics. But, no attempt was made to
  arrive at an optimised value for the minimum number of layers. Thus, 
  $n$ and I$_{0}$ are calculated from the fitted events, which are having hits in a
  minimum of 6 layers.
  
\item Similarly, the $n$ and I$_{0}$ are calculated from muons with hits in minimum of 8 layers. 
  
\item Random noise is also simulated to reproduce MC hit pattern with data, but the matching 
  was not perfect due to limited statistics in data. The effect due to possible incorrect modelling of
  noise was tested without including random noise hits in MC.
\item It was assumed that the efficiency and performance of the detector remains the same 
  during whole data taking period. But, that may not be the case. To see the time dependent
  performance of detector, experimental data are divided according to time into two sets. 
  The values of $n$ and $I_{0}$ are calculated separately for the first and the second data sets. 
\item As mentioned earlier, detector performance may not remain the same and may vary
  with time. Also, there are uncertainties in the estimation of efficiencies.
  Thus, independent data sets are used to obtain a different matrix of correlated and 
  uncorrelated efficiencies. The
  efficiency map for L0, L1, L2, L3, L4 and L5 is replaced by a new efficiency map, 
  which is calculated using the data collected with Hardware Trigger 
  L9.L10.L11 (X-plane). The efficiency map for L6, L7, L8, L9, L10 and L11 is replaced 
  by a new efficiency map, which is calculated with data collected using 
  Hardware Trigger L0.L1.L1.L3 (X-plane).
\item A complete description of different detector materials is included in the
  GEANT4 geometry description. The effect of any mismatch of material description
  in the GEANT4 geometry, the thickness of concrete above the detector is increased 
  by 20$\%$, which certainly changes the muon flux spectrum at low energies.
  
\item To see the effect of multiple scattering, the density of the aluminium tray 
  which holds the RPC is decreased by 10$\%$. This also takes care of the uncertainty of
  modelling materials within the RPC gaps.
\item The muon flux spectrum shows a peak near 1.0 GeV, which is obtained from CORSIKA simulation.
  But an experimental data \cite{kremer} predicts different energy spectrum, where peak is near 0.5 GeV and 
  slope of the energy spectrum at high energies ($dN/dE \propto E^{-slope}$) is 2.638
  (in CORSIKA, the slope is 2.624).  The energy spectrum is changed in MC, where energy spectrum 
  from \cite{kremer} was used.
\item Similarly another energy spectrum from \cite{allkofer2} was used, which has peak near 0.5 GeV and
  the slope at high energy is 2.78. 
\item The trigger efficiency maps are calculated only when muon hits are present
  anywhere in a layer. Another map was built, where efficiency is calculated only when 
  a hit is observed within 3cm of muon trajectory.     
\end{enumerate}   

\begin{table}[h]
  \center  
  \begin{tabular}{|l|c|c|c|c|}
    \hline
    &  \multicolumn{2}{c|}{$n$-value} & \multicolumn{2}{c|}{I$_{0}(\times 10^{-3}$ $cm^{-2} s^{-1} sr^{-1}$)} \\
    \hline
    Condition &  $n$ $\pm$ $\sigma(stat)$& Relative  & I$_{0}\pm \sigma(stat) \pm \sigma(syst \dag)$&  Relative \\                                                               &     &  change ($\%$) &      & change ($\%$) \\          
    \hline
    Nominal & 2.001 $\pm$ 0.039 & &7.00691 $\pm$ 0.00184 $\pm$ 0.07009 &  \\
    1	 &  2.081 $\pm$	 0.030  & + 4.0 &7.13845 $\pm$ 0.00187  $\pm$ 0.07141 & + 1.9\\
    2	 &  2.121 $\pm$  0.060 & + 6.0 &7.20532 $\pm$ 0.00189 $\pm$ 0.07207  & -2.8 \\                       
    3  & 2.019 $\pm$	0.034 &	+ 1.0 &7.02583 $\pm$ 0.00185 $\pm$ 0.07028 & + 0.3 \\            
    4 (set-1)  & 2.025 $\pm$	 0.040 & + 1.2 &7.16548 $\pm$ 0.00261 $\pm$ 0.07168 & + 2.2\\            
    4 (set-2)  & 1.973 $\pm$	0.039 & - 1.4 &6.83805 $\pm$ 0.00259  $\pm$ 0.06840 & - 2.4\\            
    5  & 1.968 $\pm$ 	0.062 &	- 1.6 &6.94597 $\pm$ 0.00183 $\pm$ 0.06948 & - 0.9\\	             
    6  & 1.972 $\pm$	0.055 & - 1.4 &6.94880$\pm$ 0.00183 $\pm$ 0.06951 & - 0.8 \\            
    7  & 1.983 $\pm$	0.055 &	- 1.0 &6.97133 $\pm$ 0.00183  $\pm$ 0.06973 & - 0.5 \\ 
    8  & 1.908 $\pm$	0.038 &	- 4.8 &7.07521$\pm$ 0.00186 $\pm$ 0.07077 & + 1.0\\             
    9 & 1.929 $\pm$	0.039 &	- 3.0 &7.15259 $\pm$ 0.00188 $\pm$  0.07155 & + 2.0\\    
    10 & 1.964 $\pm$	0.057 &	- 1.8 &7.42945  $\pm$ 0.00195 $\pm$  0.07432 & + 6.0\\                       
    \hline
  \end{tabular}\\
  $\dag$- Systematic uncertainties from $\omega$ and $\epsilon_{Daq}$.
  \caption{The value of $n$ and I$_{0}$ for different sources of systematic 
    uncertainties.}
  \label{tab:nvalue}
\end{table}
The final results with the all systematic errors are, $n$ = 2.00 $\pm$ 0.04(stat) $\pm$ 0.14(syst) and\\ $I_{0}$ = (7.0069 $\pm $0.0018(stat) $\pm$ 0.5261(syst)) $\times 10^{-3}$  $cm^{-2} s^{-1} sr^{-1}$.

\section{Comparison of results with other experiments}
\label{sec:compr}  
The angular spectra of cosmic muons are different for various locations on earth depending on its geomagnetic latitude, longitude and the altitude which have been extensively studied by various experiments\cite{greisen, judge, fukui, gokhale, karmakar, sinha, spal, allkofer1}. A comparison of the results in the current study with that of these various experiments is presented in table \ref{tab:compare}. The reults of \cite{greisen, judge, fukui, gokhale, sinha, allkofer1} suggest that the muon flux is expected to decrease with geomagnetic latitutde as moving nearer to the equator.  This can also be inferred from the variation of geomagnetic cut off rigidities which are observed to be higher at places near the equator. The result from current study, agrees with this phenomenon. In comparison with reference \cite{spal}, the present result shows a lower value of $n$ and higher value of $I_0$. The $n$ value from the current experiment is comparable with other results. 

\begin{table}[h]
  \center  
  \begin{tabular}{|l|l|l|l|l|l|l|}
    \hline
    Authors & Geomag. & Geomag. & Altitude & Muon.  & $n$ value & Integral flux \\
    & Lat.  & P${_c}$(GV) & (m)  & Mom &   &($\times 10^{-3}$ \\
    & ($^\circ$N)  &  &  & (GeV/c) &  & $cm^{-2} s^{-1} sr^{-1}$)\\  
    \hline
    Crookes and Rastin \cite{crookes} & 53 & 2.2  & 40 & $\geq$0.35 & 2.16 $\pm$ 0.01 & 9.13 $\pm$ 0.12\\
    Greisen \cite{greisen} \cite{rossi}   & 54 & 1.5 & 259 &$\geq$ 0.33 & 2.1    & 8.2 $\pm$ 0.1 \\
    Judge and Nash \cite{judge} & 53 & -- & S.L &$\geq$ 0.7 & 1.96 $\pm$ 0.22 & -- \\
    Fukui et al. \cite{fukui} & 24 & 12.6 & S.L & $\geq$ 0.34 & -- &  7.35 $\pm$ 0.2\\
    Gokhale \cite{gokhale} & 19 & -- &124 &$\geq$ 0.27  & - & 7.55 $\pm$ 0.1 \\  
    Karmakar et al \cite{karmakar} & 16 & 15.0 & 122 & $\geq$0.353 & 2.2 & 8.99 $\pm$ 0.05 \\       
    Sinha and Basu \cite{sinha} & 12 & 16.5 & 30 & $\geq$ 0.27 & -- & 7.3 $\pm$ 0.2 \\    
    S.Pal \cite{spal} & 10.61 & 16 & S.L &$\geq$ 0.280 & 2.15 $\pm$ 0.01 & 6.217 $\pm$ 0.005\\   
    Allkofer et al. \cite{allkofer1} & 9 & 14.1 & S.L &  $\geq$ 0.32  & --  & 7.25 $\pm$ 0.1\\
    $Present$ $data$ & 1.44 & 17.6  &  160 & $\geq$ 0.11 & 2.00  $\pm$ 0.04(stat) & 7.007 $\pm$ 0.002(stat)\\
    &		&		&	 &             & $\pm$ 0.14(syst)&  $\pm$ 0.526(syst)\\             		
    \hline
  \end{tabular}
  \caption{Comparison of vertical muon flux with other experiments.}
  \label{tab:compare}
\end{table}

\section{Conclusions}
\label{sec:conclu}
The angular distribution of cosmic ray muons was studied using the 2\,m\,$\times$\,2\,m RPC stack at IICHEP, Madurai. The various parameters to understand detector performance were recorded and incorporated in the MC simulation based on GEANT4. The integral exponent and integral intensity of vertical muons were estimated using the data and MC and found to be comparable with the other experimental results. The various detector parameters like position resolution, timing resolution, muon tracking efficiencies etc., observed in this RPC stack, will also be inputs to the physics simulation of the ICAL detector. A 10 layer RPC stack inside 1.5\,T magnetic field has been proposed where the muon momentum can also be estimated with some accuracy. These muon momentum spectrum, angular distribution and rate will be inputs to the neutrino event generator to obtain a better estimation of neutrino flux at INO site.

\acknowledgments{
We would like to thank Prof.D.Indumathi for many useful suggestions during this work. We acknowledge crucial contributions by A.Bhatt, S.D.Kalmani, S.Mondal, P.Nagaraj, Pathaleswar, K.C.Ravindran, M.N.Saraf, R.R.Shinde, Dipankar Sil, S.H.Thoker, S.S.Upadhya, P.Verma, E.Yuvaraj, S.R.Joshi, Darshana Koli, S.Chavan, N.Sivaramakrishnan, B.Rajeswaran, M.Sc students from Central University (Gulbarga), M.Sc students from Madurai Kamaraj University and Research Scholars from Arul Anandar College in setting up the detector, electronics and the DAQ systems.}

\end{document}